\documentclass[doublecol]{epl2}

\title{Disassortative mixing in online social networks}
\shorttitle{} 

\author{Hai-Bo Hu \and Xiao-Fan Wang}
\shortauthor{}

\institute{
  \inst{1} Complex Networks and Control Lab, Shanghai Jiao Tong University, Shanghai 200240, China
}

\abstract{ The conventional wisdom is that social networks exhibit
an assortative mixing pattern, whereas biological and technological
networks show a disassortative mixing pattern. However the recent
research on the online social networks (OSN) modifies the
wide-spread belief and many OSNs show a disassortative or neutral
mixing feature. Especially we found that an OSN, \emph{Wealink},
underwent a transition from degree assortativity characteristic of
real social networks to degree disassortativity characteristic of
many OSNs, and the transition can be reasonably elucidated by a
simple network model we propose. The relations among network
assortativity, clustering and modularity are also discussed in the
paper.}

\begin{document}

\maketitle
A social network consists of all the people-friends,
family and others-with whom one shares a social relationship, say
friendship, commerce, or others. Traditional social network study
can date back about half a century, focusing on interpersonal
interactions in small groups, not structures of large and extensive
networks due to the difficulty in obtaining large data sets [1]. The
advent of modern database technology has greatly stimulated the
statistical analysis of social networks. Novel network structures of
human societies have been revealed.

Now the WWW is undergoing a landmark revolution from the traditional
Web 1.0 to Web 2.0 characterized by social collaborative
technologies. As a fast growing business, many social networking
sites (SNS) have emerged in the Internet. The OSNs, constructed from
the SNSs and embedded in Cyberspace, have attracted attentions of
researchers from different disciplines, examples of which include
\emph{MySpace} [2], \emph{Facebook} [3], \emph{Pussokram} [4], etc.
SNSs provide an online private space for individuals and tools for
interacting with other people in the Internet. Thus the statistics
and dynamics of these OSNs are of tremendous importance to
researchers interested in human behaviors [5].

\section{Assortativity coefficient of social networks}
A structural metric of great interest in the research of social
networks, which characterizes the degree similarity of adjacent
nodes, is the degree-degree correlation, that is ``who is connected
to who?" The correlation is characterized by the assortativity $r$
and defined as the Pearson correlation coefficient $r = {{\left(
{\left\langle {ij} \right\rangle  - \left\langle i \right\rangle
\left\langle j \right\rangle } \right)} \mathord{\left/
 {\vphantom {{\left( {\left\langle {ij} \right\rangle  - \left\langle i \right\rangle \left\langle j \right\rangle } \right)} {\left( {\left\langle {i^2 } \right\rangle  - \left\langle i \right\rangle ^2 } \right)}}} \right.
 \kern-\nulldelimiterspace} {\left( {\left\langle {i^2 } \right\rangle  - \left\langle i \right\rangle ^2 } \right)}}$,
 where $i$ and $j$ are the remaining degrees at the two ends of an edge and
 the $\left\langle  \cdot  \right\rangle $ notation represents the average over all links [6].
 If a network's assorativity coefficient is negative, a hub tends to be connected to non-hubs, and vice versa.
 When $r>0$, we call the network to have an assortative mixing pattern, and
 when $r<0$, disassortative mixing. An uncorrelated network exhibits the neutral degree-mixing pattern
 whose $r=0$. While there is little systematic study of assortativity, there is a popular hypothesis that positive assortativity
 is a property of many socially generated networks, and contrasts with the opposite relationship that is more prevalent in technological and biological networks [7].

However recent extensive research on the OSNs provides many concrete
counterexamples to the prevailing view. The assortativity
coefficients for OSNs and real social networks are displayed in
Table 1. It is noteworthy that many OSNs show disassortative or
neutral mixing feature, which is in stark contrast to the
significant assortative mixing for scientific, actor, and business
collaboration networks.

\begin{table*}
\caption{Degree assortativity coefficients of OSNs and real-life
social networks. $N$ indicates the number of nodes, $r$ degree
assortativity coefficient, and percentage in parenthesis sampling
ratio.}
\begin{center}
\begin{tabular}{lllll}
\hline Type  & Network & $N$ & $r$ & References\\
\hline
Online social network  & Cyworld & 12, 048, 186 & -0.13 & [2]\\
$$  & nioki  & 50, 259 & -0.13  & [4]\\
$$ & All contacts in pussokram  & 29, 341  & -0.05 &  [4]\\
$$ & Messages in pussokram &  21, 545 & -0.06 &  [4]\\
$$ & Friends in pussokram  &  14, 278 & -0.04 &  [4]\\
$$ & Flirts in pussokram & 8, 186 & -0.12 &  [4]\\
$$ & MySpace & 100, 000 (~0.08\%)  & 0.02  &  [2] \\
$$ & orkut &  100, 000 (~0.30\%) &  0.31  &  [2] \\
$$ & Xiaonei & 396, 836  &  -0.0036 & [8] \\
$$ & Gnutella P2P(SN 6) & 191, 679  &  -0.109 & [9]\\
$$ & Flickr & 1, 846, 198 (26.9\%) & 0.202  & [10] \\
$$ & LiveJournal & 5, 284, 457 (95.4\%) & 0.179 &  [10] \\
$$ & YouTube & 1, 157, 827 & -0.033 & [10] \\
$$ & mixi  &  360, 802  &  0.1215 & [11] \\
\hline
Real social network & ArXiv coauthorship & 52, 909 & 0.36 &
[12]\\
$$ & Cond-mat coauthorship &  16, 264 & 0.18  &  [12]\\
$$ & Mathematics coauthorship  &  253, 339  &  0.12  &  [6]\\
$$ & Neuroscience coauthorship  & 205, 202  &  0.60  &  [13]\\
$$ & Biology coauthorship  &  1, 520, 251 & 0.13  &  [12]\\
$$ & Film actor collaboration  &  449, 913  &  0.21  &  [6]\\
$$ & TV series actor collaboration &  79, 663 & 0.53  &  [14]\\
$$ & Company directors  & 7, 673 & 0.28  &  [6]\\
\hline
\end{tabular}
\end{center}
\end{table*}

The origins of obvious degree assorativity for real-world social
networks are miscellaneous. From the perspective of sociology and
psychology, in real life everyone would like to have intercourse
with elites in a society; however the elites would rather
communicate with the people with the same social status as theirs,
which may lead to the assortative mixing pattern in the real-world
social networks. For professional collaborations, such as
scientific, actor, and business collaborations, the already big
names preferably collaborate with other big names for success,
reputation, influence and status. As indicated by Holme et al. [4],
assortative mixing may be significant only to interaction in
competitive areas. Another origin of degree assortativity in
professional collaboration networks is the unsubstitutability for
collaborators, which is usually decided by the similar research
interests for scholars, act styles for actors or trade backgrounds
for companies. Besides it appears that some of the degree
correlation in real social networks could have real organizational
origins. Generally the networks of collaborations between academics,
actors, and businessmen are affiliation networks, in which people
are connected together by membership of common groups (authors of a
paper, actors in a film, researchers in a lab, etc.) [15].

OSNs differ from the real-life social networks in these regards.
They break the invisible boundary among different communities or
social estates in a society. In the virtual world elites will not
refuse connections from anyone because they know that more
connections show others they are elites. And unlike in real life,
these links are not costly. Relationships in the real world have to
be maintained and this requires continual effort. The basic
difference could be the deciding difference between virtual and real
world. Understanding the process, the generative mechanism, will
supply a substantial comprehension of the formation and evolution of
online virtual communities.

\section{Assortativity transition of \emph{Wealink}}
In the following we will focus on a virtual community -
\emph{Wealink} [16], which is one of the largest OSNs in China at
present and whose users are mostly professionals, typically
businessmen and office clerks. Each registered user of the SNS has a
profile, including his/her list of friends. If we view the users as
nodes $V$ and friend relationships as edges $E$, an undirected
social network $G(V, E)$ can be constructed from \emph{Wealink}. For
privacy reasons, the data, logged from 0:00:00 h on 11 May 2005 (the
inception day for the SNS) to 15:23:42 h on 22 Aug 2007, include
only each user's ID and list of contacts, and the establishment time
for each friend relationship. The OSN is a dynamical evolving one
with the new users joining in the community and new connections
established between users.

We extract 27 snapshots of \emph{Wealink} with an interval of one
month from 11 Jun 2005 to 11 Aug 2007 and investigate the evolution
of the network. Generally it is thought that real-world networks,
man-made or naturally occurring, always belong to the same type over
time, either assortative or disassortative. However as shown in Fig.
1, the \emph{Wealink} underwent a transition from the initial
assortativity characteristic of real social networks to subsequent
disassortativity characteristic of many OSNs. To the best of our
knowledge, this is the first real-world network observed which
possesses the intriguing feature. For other SNSs, such as
\emph{Pussokram}, the assortativity coefficient of its guest book,
friend and flirt networks behaves similar to that of \emph{Wealink}
over time, however all the OSNs in \emph{Pussokram} are
disassortative and no assortativity-disassortativity transition
occurs. There could be different evolving mechanisms for real-life
and virtual social networks. A reasonable conjecture is that often
the friendship relations in the beginning OSN are based on real-life
interpersonal ones, that is \emph{Wealink} users were linking to the
other users who are their friends in the real world. In this case
the OSN directly inherits the assortative structure of the
underlying real-life social network. However at the later stage many
online users of low degrees may preferentially establish connections
with the network elites of high degrees, resulting in the
disassortative mixing. As shown in Table 1, some OSNs have
assortative mixing, however we cannot arbitrarily affirm that these
networks are all in the initial stage of evolution and reflect only
real-life interpersonal relationships. Obviously any OSN is the
superposition of online and real-life interpersonal relationships
and the different weights of both relations can regulate the
assortativity coefficient of OSNs. SNSs of differing scopes,
functions or purposes have distinct evolution pattern for the
weights of online and real-life relations. Some OSNs can always be
assortative or disassortative over time while few can undergo the
assortativity transition.

\begin{figure}
  \centerline{\includegraphics[height=2in]{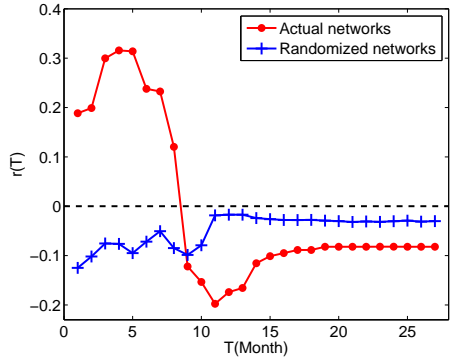}}
  \caption{Evolution of degree assortativity $r(T)$ for the \emph{Wealink} and its randomized version.}
\end{figure}

Fig. 1 also shows the comparison of $r$ between actual networks and
reshuffled ones obtained by random degree-preserving rewiring of the
original networks [17]. The randomized networks show disassortative
or nearly neutral mixing feature, and all $r$'s are less than 0 and
no transition appears. The comparison shows that the \emph{Wealink}
is strongly degree assortative at the beginning stage and
disassortative at the later stage, suggesting that individuals
indeed draw their partners from the users with degrees similar
(beginning) or dissimilar (later) to theirs far more often than one
would expect on the basis of pure chance. Fig. 2 shows the
degree-degree correlation of \emph{Wealink} on 11 Oct 2005 and 11
Apr 2006. Overall, $\left\langle {k_{{\rm{nn}}} (k)} \right\rangle $
is an increasing function of $k$ on 11 Oct 2005 and decreasing
function of $k$ on 11 Apr 2006, indicating degree assortativity and
disassortativity respectively. This can be validated by Fig. 1 and
reveals the assortativity-disassortativity transition from a
different point of view. With the increase of $k$, in Fig. 2 the
$\left\langle {k_{{\rm{nn}}} (k)} \right\rangle $ is almost flat for
the randomized networks, suggesting nearly neutral mixing.

\begin{figure}
  \centerline{\includegraphics[height=2in]{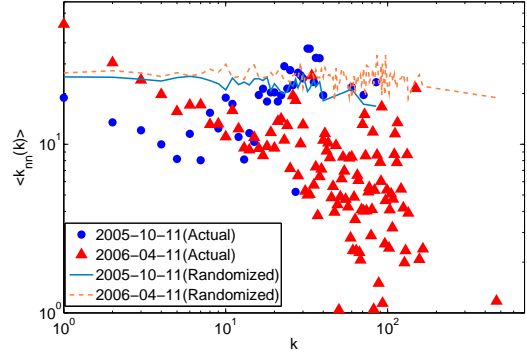}}
  \caption{$\left\langle {k_{{\rm{nn}}} (k)} \right\rangle $ versus $k$ at different time for the \emph{Wealink} and its randomized version.}
\end{figure}

Degree mixing pattern has profound effect on network structure and
behaviors, such as resistance to attacks [18], percolation [19],
epidemic spreading [20], synchronization [21] and cooperation in
games [22]. Online human interactions are driven by and can change
social conventions. A major area of web science is to explore how a
small technical innovation can launch a large social phenomenon. For
the spreading phenomena in online communities, such as diffusion of
opinions, technical innovations or gossip, one can expect the things
to be spread to a larger segment of the population in disassortative
networks than in assortative ones. Everything has two sides. The
formation of online communities facilitates human interactions and
information share in the Internet and speeds up the diffusion of
good ideas and opinions, and if exploited correctly, the OSNs can
also be a powerful medium for gauging the impact of a political
initiative or the likely success of a product launch; however at the
same time they also facilitate the spreading of vicious gossip.

A simple model based on the work of Catanzaro et al. [23] can
reproduce the assortativity transition. Starting with a small random
connected network, at every time step:

(1) With probability $p$ a new node is added into the network, and
it is linked to an old node $i$ by Barab\'{a}si-Albert (BA)
preferential attachment rule [24], i.e. $pk_i /\sum\nolimits_{j =
1}^N {k_j } $, where $k_i$ is the degree of node $i$ and $N$ is the
number of nodes.

(2) With probability $(1-p)$ a new edge is added between two
existing nodes (self-loops and multiple links are prohibited), which
are chosen based on their degrees $k_i$, $k_j$. The probability of
selecting the first node with degree $k_i$ is $P_1 (k_i )$, and the
second node $P_2 (k_j |k_i )$. Thus the probability of adding a new
edge and connecting two old nodes is $(1 - p)P_1 (k_i )P_2 (k_j |k_i
)$. The first node is selected based on BA rule. The functional form
of $P_2 (k_j |k_i )$ can be chosen so as to favor links between
similar or different degrees. We let $P_2 (k_j |k_i ) \propto (1 -
p')f_1 (\left| {k_i  - k_j } \right|) + p'f_2 (\left| {k_i  - k_j }
\right|)$, where $f_1$ and $f_2$ are suitable decreasing functions
and increasing functions of $\left| {k_i - k_j } \right|$ with
positive range respectively, and $0 \le p' \le 1$ governs the weight
of degree assortativity. By tuning $p'$ from 0 to 1, the resulting
network will undergo a gradual structural transition from
assortativity to disassortativity.

The $p$ modulates the relative role of growth and mixing. Generally
$p<0.5$ because the mixing is often more frequent than growth in
social networks, i.e. the mean life of a node (a human or
professional life) is longer than that of an edge (a business or
social relation). Let $P_2 (k_j |k_i ) \propto (1 - p'){\mathop{\rm
e}\nolimits} ^{ - \left| {k_i  - k_j } \right|}  + p'{\mathop{\rm
e}\nolimits} ^{\left| {k_i - k_j } \right|}$, and $p=0.1$, 0.2 and
0.3, we generate three evolving networks, whose assortativity
coefficients $r$ are shown in Fig. 3. Indeed the model cannot
reproduce the realistic and intricate evolving process of
\emph{Wealink} in many aspects since the real growth of the network
will inevitably be affected by various exogenous and endogenous
factors, however the qualitative agreement between Fig. 1 and Fig. 3
shows that the model could reproduce the macroscopical property of
assortativity-disassortativity transition, including the
microscopical mechanism of growth and mixing, and thus capture the
basic aspect governing the evolution of $r$ of the network.

\begin{figure}
  \centerline{\includegraphics[height=2in]{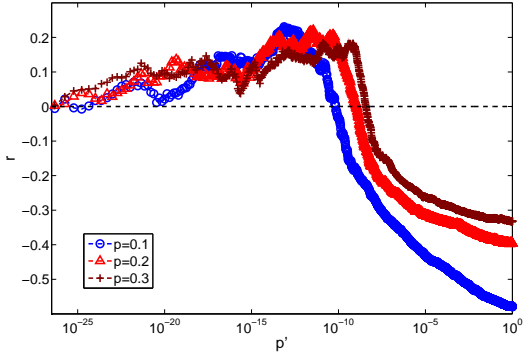}}
  \caption{Simulation result of $r$ with the increase of $p'$. ${\mathop{\rm e}\nolimits} ^{\left| {k_i  - k_j } \right|}$ has a much stronger effect on
  $P_2 (k_j |k_i )$ than ${\mathop{\rm e}\nolimits} ^{ - \left| {k_i  - k_j } \right|}$, thus $r>0$ occurs only for extremely small $p'$.}
\end{figure}

In the above model, we assume that when a new node is added into the
network, it is attached to an old node by BA rule, and when a new
edge is added between two old nodes, the first node of creating link
request also is selected based on BA rule. The linear preference can
be validated by real network data [25]. Let $k_i$ be the degree of
user $i$. The probability that user $i$ with degree $k_i$ is chosen
can be expressed as $\prod (k_i ) = k_i^\beta  /\sum\nolimits_j
{k_j^\beta  } $, where $\beta$ is a constant. We can compute the
probability $\prod (k)$ that an old user of degree $k$ is chosen,
and it is normalized by the number of users of degree $k$ that exist
just before this step: $\prod (k) = {{\sum\nolimits_t {\left[ {e_t =
v \wedge k_v (t - 1) = k} \right]} } \mathord{\left/
 {\vphantom {{\sum\nolimits_t {\left[ {e_t  = v \wedge k_v (t - 1) = k} \right]} } {\sum\nolimits_t {\left| {\left\{ {u:k_u (t - 1) = k} \right\}} \right|} }}} \right.
 \kern-\nulldelimiterspace} {\sum\nolimits_t {\left| {\left\{ {u:k_u (t - 1) = k} \right\}} \right|} }} \sim k^\beta
 $, where $e_t  = v \wedge k_v (t - 1) = k$ represents that at time $t$
the old user whose degree is $k$ at time $t-1$ is chosen. We use $[
\cdot ]$ to denote a predicate (take value of 1 if expression is
true, else 0). Generally $\prod (k)$ has significant fluctuations,
particular for large $k$. To reduce the noise level, instead of
$\prod (k)$ we study the cumulative function: $\kappa (k) = \int_0^k
{\prod (k)} {\rm{d}}k \sim k^{\beta  + 1} $. Fig. 4 shows how the
degree $k$ of users is related to the preference metric $\kappa$.
Approximately $\beta  \approx 1$ for both preferential attachment
and creation, which indicates that the linear preference hypothesis
is reasonable.

\begin{figure}
  \centerline{\includegraphics[height=2in]{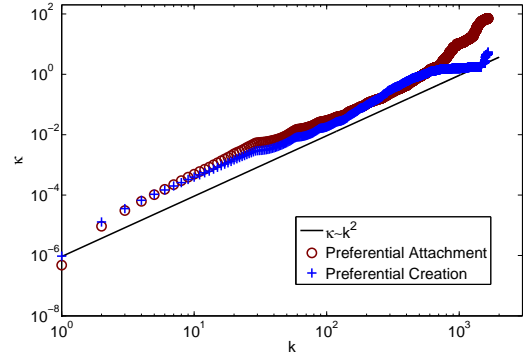}}
  \caption{Testing preference feature for \emph{Wealink}.}
\end{figure}

Fig. 5 shows the evolution of clustering coefficient $C$ [26] and
modularity $Q$ [27], in comparison with the same metric of
randomized networks. $C$ measures the strength of connections within
individual neighborhood and $Q$ measures the significance of
community feature. The $Q$ lies in $[0, 1)$ and a $Q$ value above
about 0.3 is a good indicator of significant community structure in
a network. As shown by Fig. 5, the \emph{Wealink} has significantly
higher $C$ and $Q$ than those of randomized networks and both $C$
and $Q$ of actual networks are similar in growth trend to those of
randomized ones. We find that the randomized networks still possess
large $Q$, which may result from the structural constraint of degree
sequence of original networks.

\begin{figure*}
  \centerline{\includegraphics[height=2in]{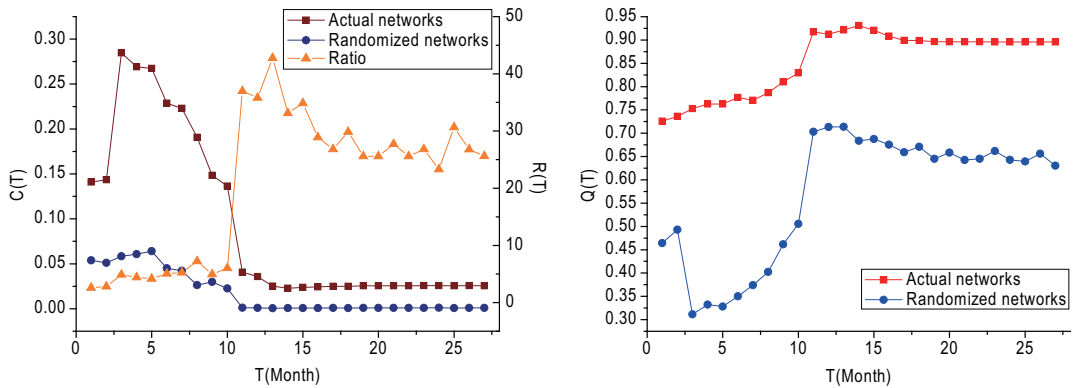}}
  \caption{Evolution of clustering coefficient $C(T)$ (left) and modularity $Q(T)$ (right)
  for the actual networks and randomized ones. The left scale of y-axis in the left panel is for clustering
  coefficient of \emph{Wealink} and its randomized version and the right scale the ratio of $C$ of actual networks to that of randomized ones.}
\end{figure*}

A social network might be divided up according to the location,
affiliation, occupation, interests, and so forth, of its members. It
is thought that clustering and assortativity in networks arise
because the vertices are divided into groups or communities [7, 28]
with a high density of edges between members of the same group, even
though the density of edges in the network as a whole may be low;
however the comparison between Fig. 1 and Fig. 5 shows that both
clustering coefficient and degree assortativity are negatively
correlative with modularity. Recent research also shows that the
$r-C$ space obtained by successively rewiring pairs of edges of
networks suggests that there exists positive correlation to some
extent between assortativity coefficient $r$ and clustering
coefficient $C$ [29], which is intuitively reasonable and is
validated empirically by Figs. 1 and 5. However it should be quite
cautious to claim that there exists specific correlation between
network metrics, and obviously under different conditions there may
be quite different conclusions. In practice different network
properties, such as modularity, clustering, assortativity,
heterogeneity, synchronizability, etc., may constrain each other, or
not be independent [30]. And the arbitrary claim about correlation
usually may lead to unsound conclusions.

\section{Summary}
In summary, we systematically study the assortativity of social
networks. We find that, compared to real social networks, OSNs show
diverse degree correlation pattern, including disassortative,
assortative or nearly neutral mixing, which implies different
evolving mechanisms for real world and virtual community. More
interesting we have found that an online community, \emph{Wealink},
underwent a transition from degree assortativity to
disassortativity, which can be reasonably interpreted by a simple
model we propose. As a rapidly developing field in interdisciplinary
research, virtual community has attracted scholars of different
backgrounds, mostly physicists and computer scientists. However the
main body in the virtual world is still persons in real world, thus
understanding the web community may also require insights from
sociology and psychology every bit as much as from physics and
computer science [31].

\acknowledgments We thank Wealink Co. for providing the network data
and Dr. Tao Zhou for the valuable comments and suggestions. We also
thank the anonymous reviewers for their constructive remarks and
suggestions which helped us to improve the quality of the manuscript
to a great extent. This work was partly supported by the NSF of PRC
under Grant No. 60674045.

\end{document}